\documentclass[preprint,prd,tightenlines,superscriptaddress]{revtex4-2}

\usepackage[dvipdfmx]{graphicx}
\usepackage{dcolumn}  
\usepackage{colordvi}
\usepackage{color}
\usepackage{epstopdf}
\usepackage{amssymb}
\usepackage{url}
\graphicspath{{ps}}
\usepackage{hyperref}
\usepackage{tabularx}
\usepackage{amsmath}
\usepackage{tabularx}
\usepackage{microtype}
\usepackage{hyperref}
\usepackage[italic]{hepnames}
\usepackage{hepparticles}
\usepackage{physics}
\usepackage{siunitx}
\usepackage{bm}
\usepackage[noabbrev, capitalise]{cleveref}

\usepackage[section]{placeins}

\input ./belle2sym.tex

\begin{document}

\def\belletwo {\it {Belle II}}

\vspace*{-3\baselineskip}
\resizebox{!}{3cm}{\includegraphics{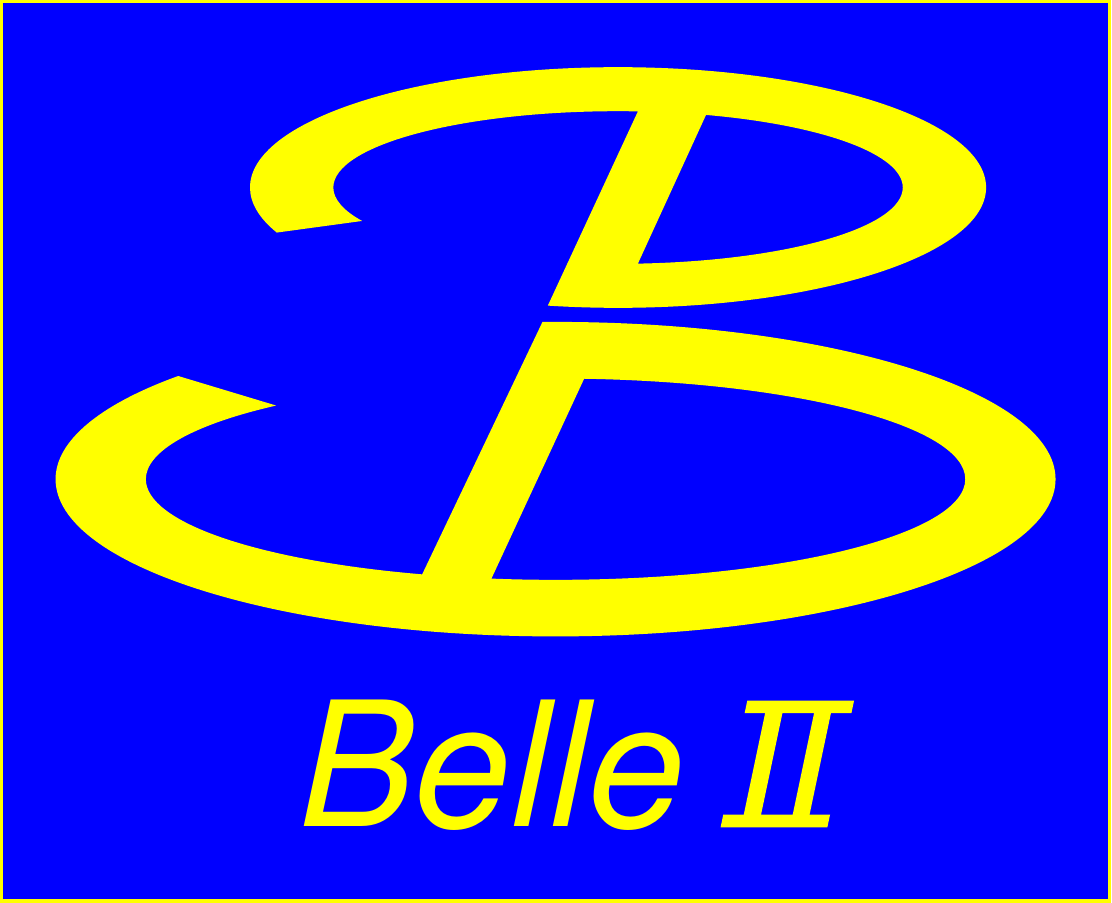}}

\vspace*{-5\baselineskip}
\begin{flushright}
BELLE2-CONF-PH-2021-037\\
\today
\end{flushright}

\title { \quad\\[0.5cm] Measurement of the  $\Bz \rightarrow \Dstarm \ellp \nul$ branching ratio and $|V_{cb}|$ with a fully reconstructed accompanying \B meson in 2019-2021 Belle II data  }


\collaboration{The Belle II Collaboration}
  \author{F. Abudin{\'e}n}
  \author{I. Adachi}
  \author{K. Adamczyk}
  \author{L. Aggarwal}
  \author{P. Ahlburg}
  \author{H. Ahmed}
  \author{J. K. Ahn}
  \author{H. Aihara}
  \author{N. Akopov}
  \author{A. Aloisio}
% ...


\begin{abstract}
	We present a measurement of the $\Bz \rightarrow \Dstarm \ellp \nul $ ($\ell=\electron,\mu$) branching ratio and of the CKM parameter $|V_{cb}|$ using signal decays accompanied by a fully reconstructed \B meson. The Belle II data set of electron-positron collisions at the \FourS resonance, corresponding to 189.3$\,$fb$^{-1}$ of integrated luminosity, is analyzed. With the Caprini-Lellouch-Neubert form factor parameterization, the parameters $\eta_{\rm EW} F(1) |V_{cb}|$ and $\rho^{2}$ are extracted, where $\eta_{\rm EW}$ is an electroweak correction, $F(1)$ is a normalization factor and $\rho^{2}$ is a form factor shape parameter. We reconstruct 516 signal decays and thereby obtain $\mathcal{B} (\Bz \rightarrow \Dstarm \ellp \nul ) =  \left(5.27 \pm 0.22~\rm{\left(stat\right)} \pm 0.38~\rm{\left(syst\right)}\right) \%$
, $\eta_{EW} F(1) |V_{cb}| \times 10^{3} =  34.6 \pm 1.8~\rm{\left(stat\right)} \pm 1.7~\rm{\left(syst\right)}$, and $\rho^{2} =   0.94   \pm 0.18~\rm{\left(stat\right)} \pm 0.11~\rm{\left(syst\right)}$.

\keywords{Belle II, ...}
\end{abstract}

\pacs{}

\maketitle

{\renewcommand{\thefootnote}{\fnsymbol{footnote}}}
\setcounter{footnote}{0}

\section{Introduction}

A precise understanding of $\Bz \rightarrow \Dstarm \ellp \nul$ decays is important for future measurements of $R(\Dstar)=\mathcal{B} (B \rightarrow \Dstar \tau \nu) / \mathcal{B}(B \rightarrow \Dstar \ell \nu)$~\cite{HFLAV:2019otj,Fajfer:2012vx} 
and of the magnitude of the Cabibbo-Kobayashi-Maskawa matrix element $V_{cb}$~\cite{cite:C,cite:KM}, where persistent tensions exist between inclusive $\B \rightarrow X_{c} \ell \nu$ and exclusive $\B \rightarrow \Dstar \ell \nu$ measurements~\cite{HFLAV:2019otj}. We study $e^{+}e^{-} \rightarrow \FourS \rightarrow$ \BzBzb events, where the decay of the accompanying \Bz or \Bzb is reconstructed in a hadronic final state using the full event interpretation algorithm (FEI)~\cite{cite:FEI} and the signal bottom meson of opposite flavor is then reconstructed in the $\Dstarpm \ell^{\pm} \nul$ final state.

\section{Belle II experiment}

Belle II~\cite{cite:bellell} is an experiment at the SuperKEKB super $B$ factory~\cite{cite:skekb}, an energy-asymmetric $e^{+}$ (4$\,$GeV) $e^{-}$ (7$\,$GeV) collider
in Tsukuba, Japan. Collision data with an integrated luminosity corresponding to 189.3$\,$fb$^{-1}$ were collected from March 2019 to July 2021 at a center-of-mass (c.m.) energy of 10.58$\,$GeV, corresponding to the mass of the \FourS resonance, as well as 18.0$\,$fb$^{-1}$ at 60$\,$MeV below the nominal c.m. energy.

The Belle II detector consists of several nested detector subsystems arranged around the beam pipe in a cylindrical geometry. The innermost subsystem is the vertex detector, which includes one or two layers of silicon pixels and four outer layers of silicon strips. Outside the silicon, the central drift-chamber reconstructs charged-particles trajectories (tracks). Outside the chamber, a Cherenkov light-imaging and time-of-propagation detectors provide charged particle identification. Further out is an electromagnatic calorimeter with CsI(Tl) crystals. A uniform 1.5$\,$T magnetic field aligned with the beam axis is provided by a superconducting solenoid. Multiple layers of scintillators and resistive plate chambers, located between the magnetic flux-return iron plates, detect $K^{0}_{L}$ and muons.

The analysis uses simulated Monte Carlo (MC) samples to determine the signal efficiency and background yields. 
These samples are generated using EvtGen~\cite{cite:evtgen} and consist of $e^{+}e^{-} \rightarrow \FourS \rightarrow \BB $ (generic) and $e^{+}e^{-} \rightarrow \qqbar$ processes, where $B$ indicates a \Bz or a \Bp meson and $q$ indicates an $u$, $d$, $c$, or $s$ quark (continuum). The latter is simulated with KKMC~\cite{cite:kkcm} and PYTHIA~\cite{cite:pythia}. The luminosity of the generic and continuum samples is 1$\,$ab$^{-1}$.
The signal is modeled using the CLN form factor parameterization~\cite{cite:cln}, and the time-integrated $\BzBzb$-mixing parameter~\cite{cite:pdg2020}.
All samples are analyzed with the basf2 framework~\cite{cite:basf2,basf2-zenodo}. In this paper, the natural system of units with $c=\hbar=1$ is used.
The inclusion of charge-conjugated decay modes is implied unless otherwise stated.

\section{Event selection}

The reconstruction begins by fully reconstructing a 
\Bz or \Bzb ($B_{\rm tag}$) in hadronic decay modes with the FEI algorithm~\cite{cite:FEI}. 
The algorithm starts by selecting candidates for stable particles, which include muons, electrons, pions, protons, kaons, and photons, from tracks and electromagnetic energy deposits in each event. Subsequently, the algorithm carries out several stages of reconstruction of intermediate particles such as \piz, 
$K^{0}_{S}$, $J/\psi$, 
$D$ and \Dstar mesons, $\Sigma$, $\Lambda$, and $\Lambda_{c}$ baryons. Intermediate particles are reconstructed in specific decay modes from combinations of stable and other intermediate particle candidates. The final stage of the algorithm reconstructs the \Bz mesons in 31 hadronic modes, using boosted decision trees (BDTs). The $B_{\rm{tag}}$ candidates are required to have a BDT classifier output greater than 0.001, a beam constrained mass $M_{\rm{bc}} = \sqrt{E_{\rm{beam}}^{2}-|\vec{p}_{B_{\rm{tag}}}|^{2}} > 5.27\,$GeV , and an energy difference $\Delta E=E_{B_{\rm{tag}}}-E_{\rm{beam}}$ in the interval [-0.15, 0.1]$\,$GeV, where $E_{\rm{beam}}$ is half of the collision energy, and $\vec{p}_{B_{\rm{tag}}}$ and $E_{B_{\rm{tag}}}$ are the momentum and energy of the $B_{\rm{tag}}$ candidate, all in the center-of-mass frame. The efficiency of the FEI is calibrated with $\B \rightarrow X \ell \nu$ decays~\cite{cite:FEI}.

\begin{figure}[!htb]
\begin{minipage}{5.0cm}
        \centering
        \includegraphics[width=5.0cm]{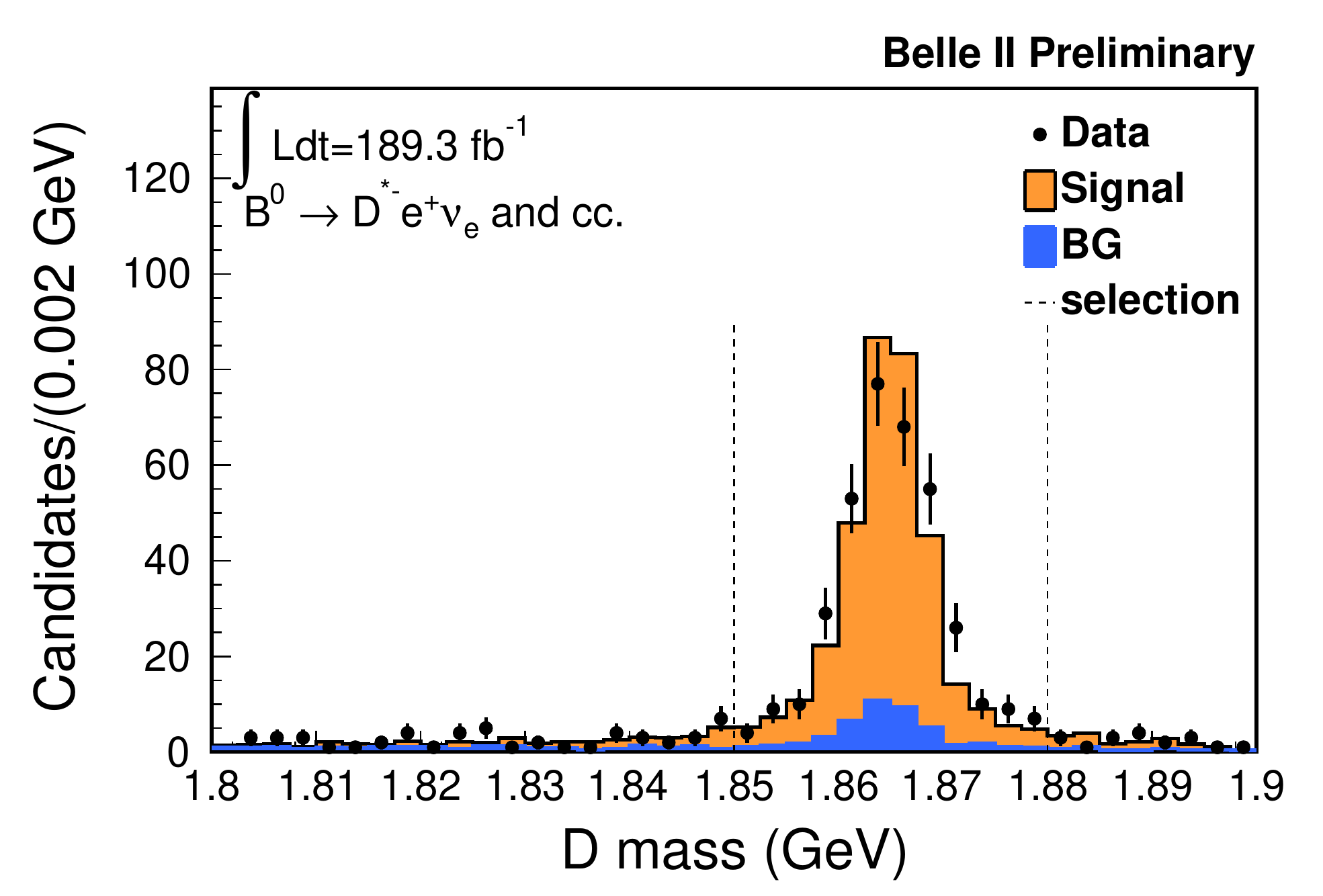}
\end{minipage}
\begin{minipage}{5.0cm}
        \centering
        \includegraphics[width=5.0cm]{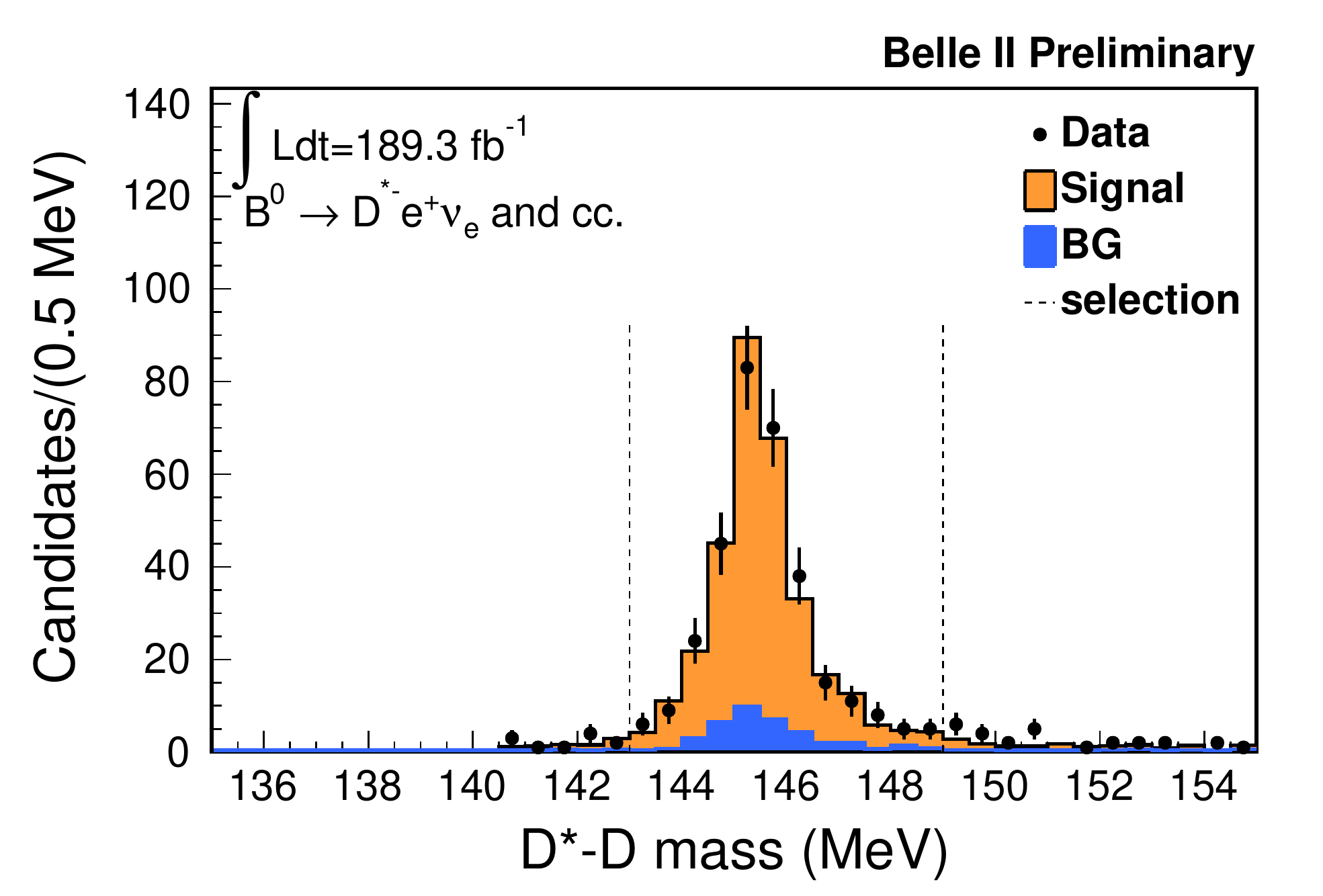}
\end{minipage}
\begin{minipage}{5.0cm}
        \centering
        \includegraphics[width=5.0cm]{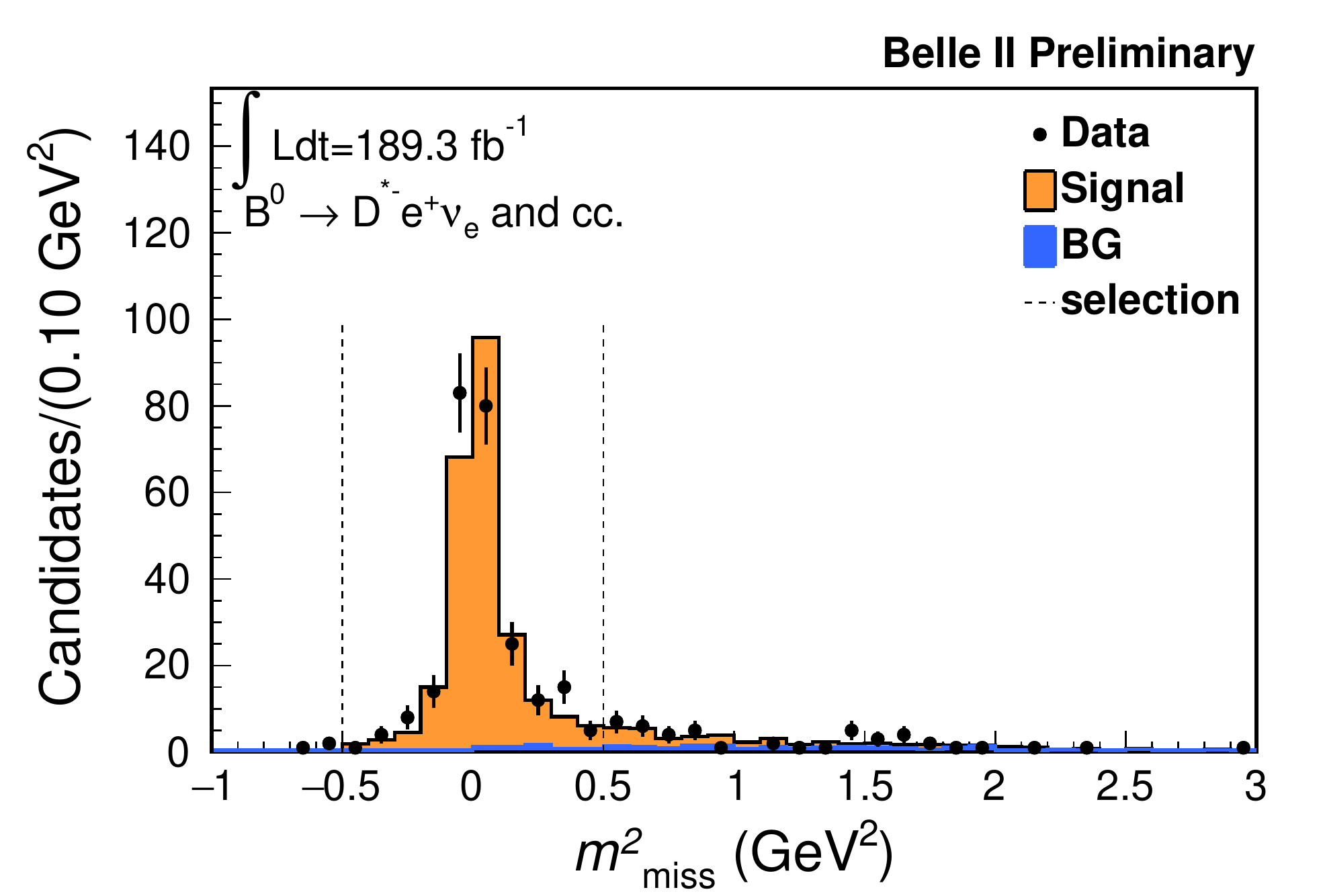}
\end{minipage}\\
\begin{minipage}{5.0cm}
        \centering
        \includegraphics[width=5.0cm]{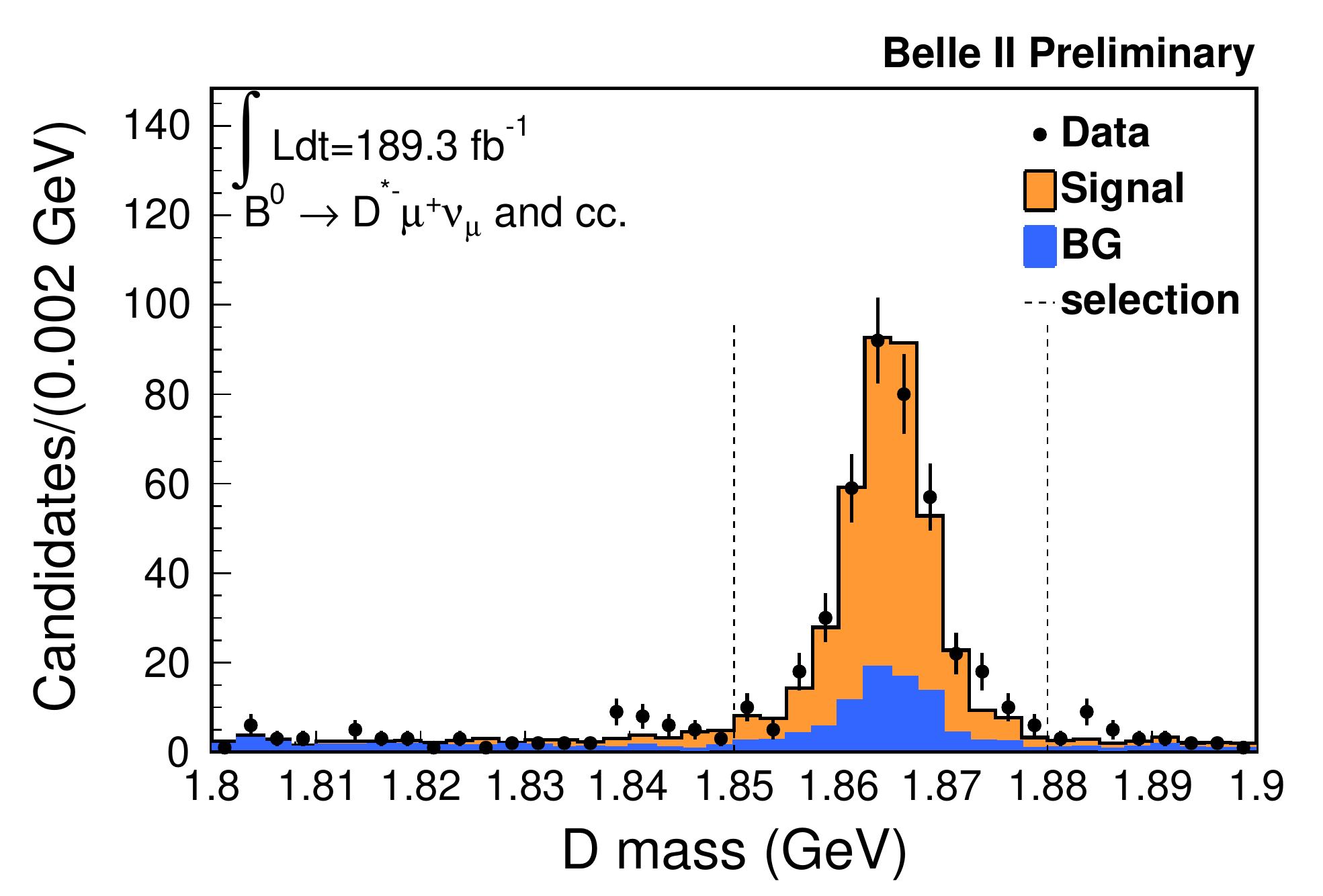}
\end{minipage}
\begin{minipage}{5.0cm}
        \centering
        \includegraphics[width=5.0cm]{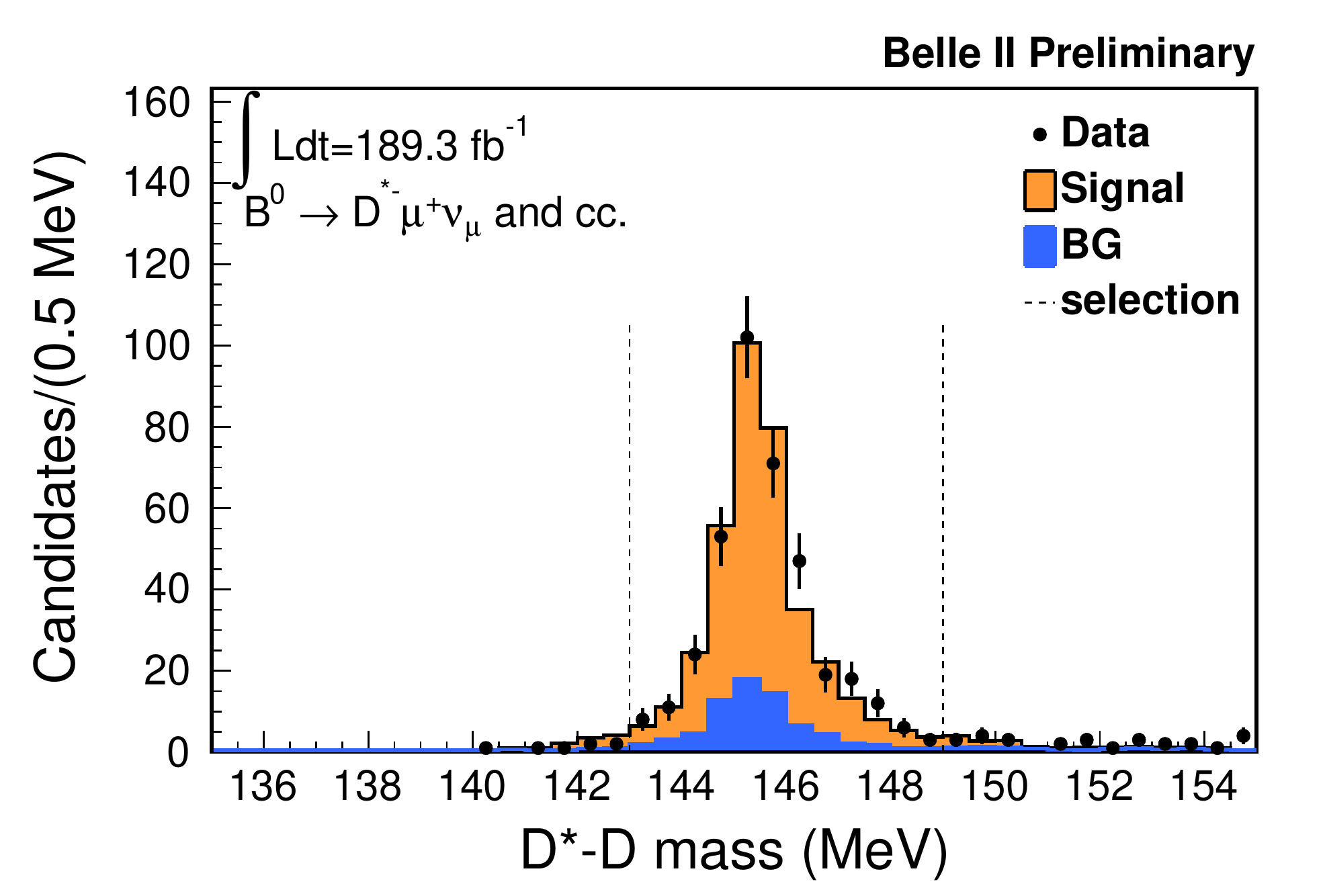}
\end{minipage}
\begin{minipage}{5.0cm}
        \centering
        \includegraphics[width=5.0cm]{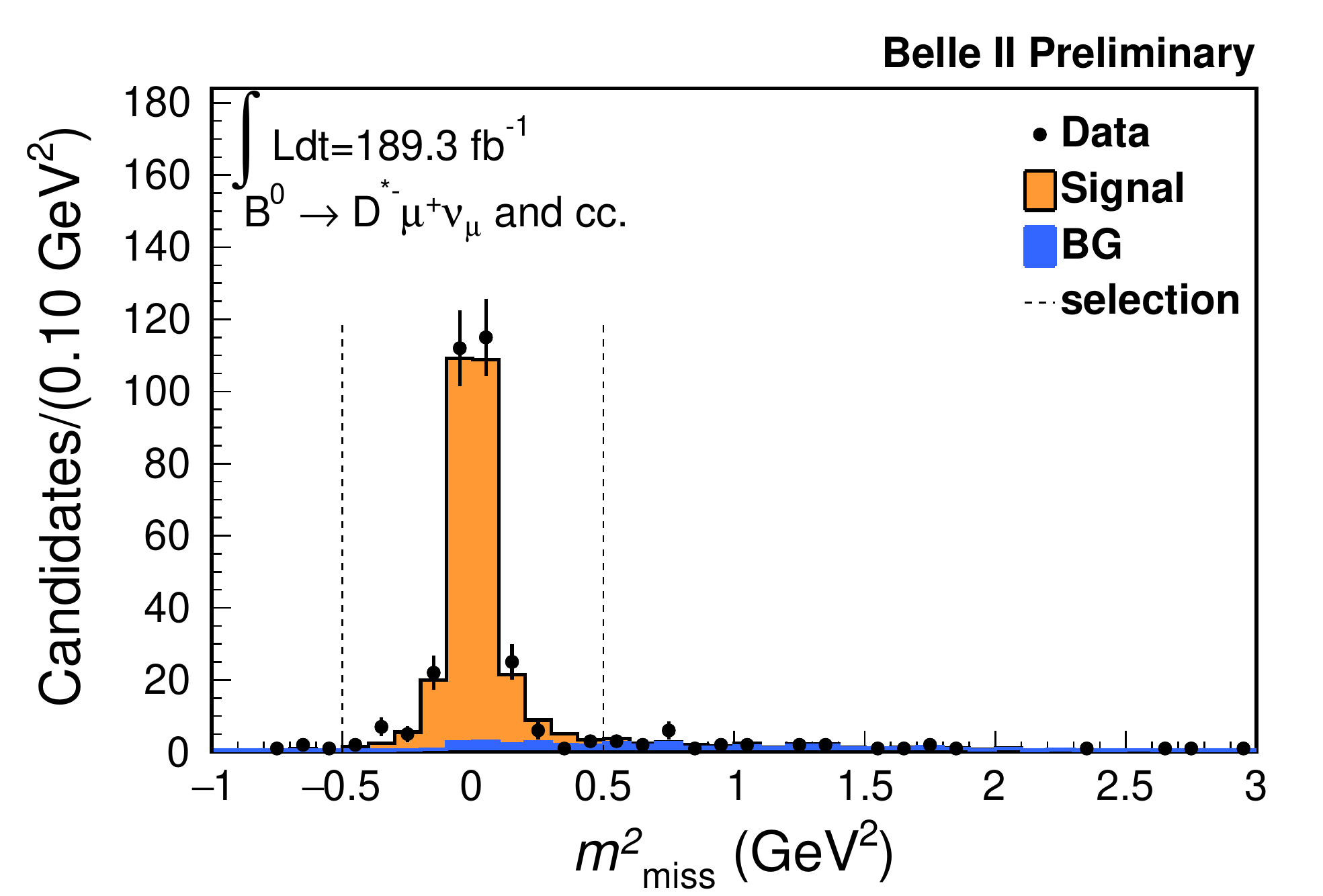}
\end{minipage}
\caption{ Distributions of $K \pi$ invariant mass (left), \Dstar - \Dz mass difference (middle), and missing neutrino mass squared (right) of $\Bz \rightarrow \Dstarm e^{+}$ $\nu_{e}$ (top) and $\Bz \rightarrow \Dstarm \mu^{+} \nu_{\mu}$ (bottom) candidates in data (points) and simulation (histograms). Vertical lines enclose the regions of the selected events. }
  \label{ fig:sel_3 }
\end{figure}

An event-level selection requires more than three tracks, between 2 and 7$\,$GeV of total energy in the electromagnetic calorimeter, and a ratio of the second to zeroth order Fox Wolfram moments, R2~\cite{PhysRevLett.41.15812}, to be smaller than 0.4.
The remaining \Bz or \Bzb meson (the signal side $B_{\rm{sig}}$) is reconstructed in its decay of
 $\Bz \rightarrow \Dstarm \ellp \nul$ ( $\Dstarm \rightarrow \pim \Dzb$, $\Dzb \rightarrow \Kp \pim$).
Charged particles are required to originate from the interaction point and have a transverse momentum greater than 0.2$\,$GeV. To identify electrons and muons, a likelihood-ratio like quantity for each particle hypothesis is calculated, which combines information from several detector subsystems.
The likelihood performance is calibrated with well-known physics processes. In addition, electron and muon momenta are required to be greater than 1.0$\,$GeV in the center-of-mass frame to reject continuum background. 
Kaon and pion candidates are combined to reconstruct $\Dzb \rightarrow \Kp \pim$ candidates whose invariant mass $\left( m \left( K \pi \right) \right)$ is required to be within the interval [1.85, 1.88]$\,$GeV. The \Dzb candidates are combined with an additional low-momentum pion to reconstruct $\Dstarm \rightarrow \pim \Dzb$
candidates restricted to the \Dstarm - \Dzb mass difference $\Delta m$ in the range [0.143, 0.149]$\,$GeV.
Subsequently, $B_{\rm{sig}}$ candidates are reconstructed by combining \Dstarm candidates with either $\ep$ or $\mup$ candidates. At least one combination of $B_{\rm{tag}}$ and $B_{\rm{sig}}$ candidates is required with no remaining tracks. The missing neutrino mass squared ($m^{2}_{\rm{miss}}= (P_{\rm{beam}}-P_{B_{\rm{tag}}} - P_{\Dstar} - P_{\ell})^{2}$, where $P$ denotes a four vector) is required to be in the range [-0.5, 0.5]$\,$GeV$^{2}$. If multiple combinations of $B_{\rm{tag}}$ and $B_{\rm{sig}}$ candidates are found in an event, the candidate with the largest BDT output for the $\B_{\rm{tag}}$ and the best $\Delta m$ for the $B_{\rm{sig}}$ is selected. 
Figure~\ref{ fig:sel_3 } shows the $m \left( K \pi \right)$, $\Delta m$, and $m_{\rm{miss}}^{2}$ distributions. 
%
The figures include data points with statistical uncertainties and histograms for simulated signal and background candidates scaled to the equivalent data luminosity. The signal yield is estimated by counting the number of selected events on data from which simulated background is subtracted. Checks based on data in the $\Delta m$ sidebands show good agreement between simulated and experimental background distributions.

\section{Measurement of branching ratio}

The branching ratio for the decay $\Bz \rightarrow \Dstarm \ellp \nul$ is estimated as

\begin{align}
\mathcal{B} \left(\Bz \rightarrow \Dstarm \ellp \nul \right)  &= \frac{ \left(N^{\rm{rec}} - N^{\rm{bg}} \right) \epsilon^{-1} }{ 4 N_{\BB} \left(1+f_{+0} \right)^{-1} \mathcal{B}\left(\Dstarm \rightarrow \pim \Dzb \right) \mathcal{B}\left(\Dzb \rightarrow \Kp \pim \right)}, \label{ eq:br0 }
\end{align}
where $N^{\rm{rec}}$ is the number of reconstructed events in data, $N^{\rm{bg}}$ is the number of reconstructed background events, $\epsilon$ is the signal reconstruction efficiency, $N_{\BB}$ is the number of produced $\BB$ pairs, and $f_{+0}$ is the ratio of the number of produced $\BpBm$ and $\BzBzb$ pairs. In Eq.~\ref{ eq:br0 }, the values corresponding to electron and muon modes are averaged. 
The values of $N^{\rm{bg}}$ and $\epsilon$ are estimated from the background and signal simulation.
The value of $N_{\BB}$ is determined using the R2 distribution after a subtraction of the continuum background using off-resonance data. The values for $f_{+0}$, $\mathcal{B}(\Dstarm \rightarrow \pim \Dzb )$, $\mathcal{B}(\Dzb \rightarrow \Kp \pim )$, and fraction of mixed and unmixed $\BzBzb$ are taken from Ref.~\cite{cite:pdg2020}. 
The input values for the branching fraction measurement are summarized in Table~\ref{ tab:br_1 }. 

\begin{table}[!htb]
	\begin{center}
		\caption{ Input values for the measurement of branching ratio with the systematic uncertainties, described in Sec.~\ref{ sec:sys }.  }
		\label{ tab:br_1 }
		\begin{tabular}{lccccccccccccccccccccccccc}
			\hline
			\hline
			Variables                       & Values \\
			\hline
			$N^{\rm{rec}}$                  &  $545$ (data)    \\
			$N^{\rm{bg}}$                   &  $29.4 \pm 11.2$        \\
			$\epsilon$                      &  $(9.55 \pm 0.67)  \times 10^{-4}$   \\   
			$N_{\BB}$                       &  $(197.17 \pm 5.72) \times 10^{6}$ \\
			$f_{+0}$                        &  $1.058  \pm 0.024$   \\
			$\chi_{d}$                      &  $0.1875 \pm 0.0017$   \\
			$\mathcal{B}(\Dstarm \rightarrow \pim \Dzb)$        &  $(67.7 \pm 0.5) \%$       \\
 			$\mathcal{B}(\Dzb \rightarrow \Kp \pim)$            &  $(3.950 \pm 0.031) \%$   \\
			\hline
			\hline
		\end{tabular}
	\end{center}
\end{table}

\section{Measurement of $|V_{cb}|$} \label{ sec:vcb }

\begin{figure}[!htb]
\begin{minipage}{8.0cm}
        \centering
        \includegraphics[width=8.0cm]{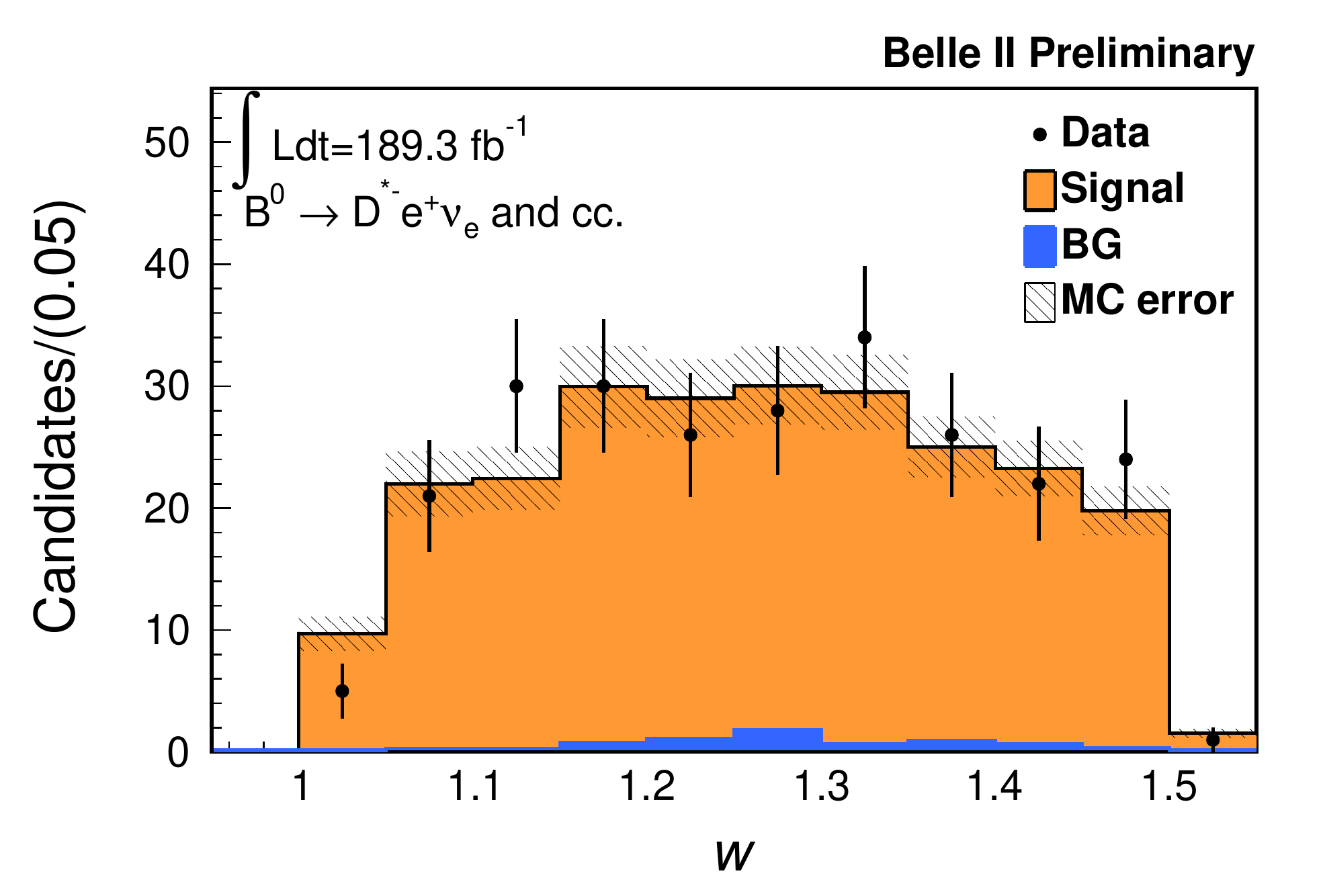}
\end{minipage}
\begin{minipage}{8.0cm}
        \centering
        \includegraphics[width=8.0cm]{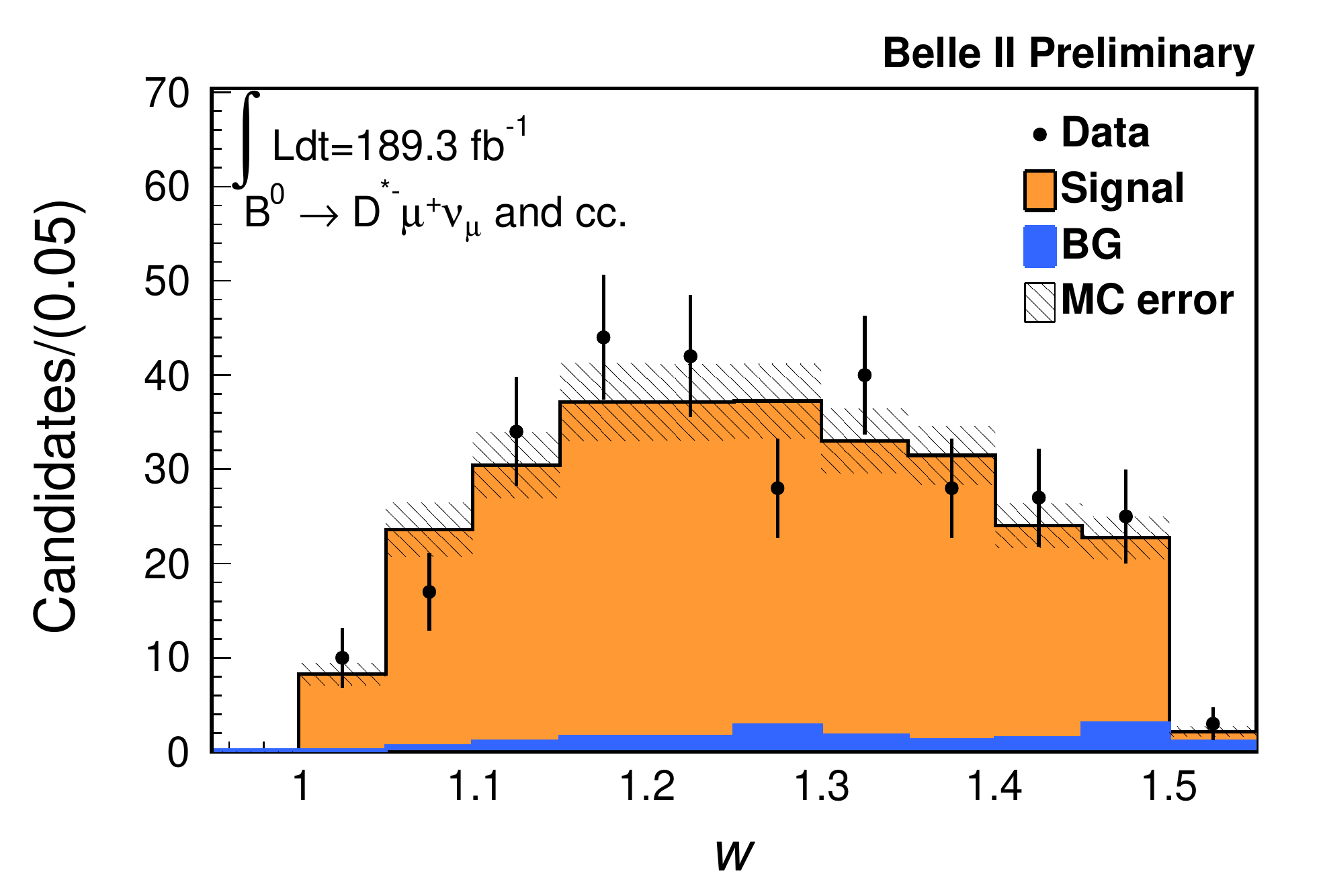}
\end{minipage}
\caption{ Distributions of $w$ for $\Bz \rightarrow \Dstarm e^{+} \nu_{e}$ (left) and $\Bz \rightarrow \Dstarm \mu^{+} \nu_{\mu}$ (right) candidates in data (points) and simulation (histogram) after the event selection. The shaded band shows the systematic uncertainty of the simulation, which is summarized in Sec.~\ref{ sec:sys }. }
  \label{ fig:sel_7 }
\end{figure}

Figure~\ref{ fig:sel_7 } shows distribution of the recoil variable

\begin{align}
w = \frac{P_{B}\cdot P_{\Dstar}}{m_{B}m_{\Dstar}} = \frac{m^{2}_{B}+m^{2}_{\Dstar}-q^{2}}{2m_{B}m_{\Dstar}},
\end{align}
where $q^{2} = (P_{\ell}+P_{\nul})^2$ and $m_{B,\Dstar}$ are the known masses of the indicated particles.
The $\Bz \rightarrow \Dstarm \ellp \nul$ decay-width differential in $w$ is as follows~\cite{cite:KM,cite:neubert}:

\begin{align}
	\frac{d\Gamma}{dw} = \frac{ \eta_{\rm EW}^{2} \color{black} G^{2}_{F}}{48\pi^{3}} m^{3}_{\Dstar}(m_{B}-m_{\Dstar})^{2}g(w)F^{2}(w)|V_{cb}|^{2}. \label{ eq:dgdw }
\end{align}
Here, $\eta_{\rm EW}$ is an electroweak correction 
(calculated to be $1.00662 \pm 0.00016$ in Ref.~\cite{cite:qcd2}). From lattice QCD, $F(1)$ is calculated as $0.906 \pm 0.004~\rm{(stat)} \pm 0.012~\rm{(syst)}$~\cite{cite:qcd2}. 
The product $g(w)F^{2}(w)$ describes the phase-space factor and the form factor, which is parameterized with $R_{1}(1), R_{2}(1), \rho^{2}$ in the CLN approach~\cite{cite:cln}.
The CKM matrix element $|V_{cb}|$ is determined by fitting the $\Delta \Gamma / \Delta w$ distribution with the form factor parameters. However, here the product $\eta_{\rm EW} F(1) |V_{cb}|$ is measured instead, in order to separate theory uncertainty of $\eta_{\rm EW}$ and $F(1)$. 
In this paper, the $R_{1}(1)$ and $R_{2}(1)$ values are taken from external measurements~\cite{HFLAV:2019otj}, as shown in Table~\ref{ tab:r1r2 }.

\begin{table}[!htb]
	\begin{center}
		\caption{ Input values for $R_{1}$(1) and $R_{2}$(1)~\cite{HFLAV:2019otj}.  }
		\label{ tab:r1r2 }
		\begin{tabular}{lccccccccccccccccccccccccc}
			\hline
			\hline

			 $R_{1}(1)$              &   1.270 $\pm$ 0.026    \\
			 $R_{2}(1)$              &   0.852 $\pm$ 0.018    \\
			 Correlation coefficient of $R_{1}(1)$ and $R_{2}(1) $&  -0.715  \\
			\hline
			\hline
		\end{tabular}
	\end{center}
\end{table}

\subsection{Unfolding method}

In order to estimate the true $w$ distribution from the observed $w$ values, an iterative unfolding method is used~\cite{cite:unfolding}. The number of signal events populating $w$ bin, $N_{i}$, is estimated from the reconstructed variables as follows:

\begin{align}
	N_{i} = \sum_{j} U_{ij} (N^{\rm{rec}}_{j}-N^{\rm{bg}}_{j}),  \label{ eq:dgdw_bp } 
\end{align}
where $i$ is the $w$ bin number. We define 10 bins in the range $[1.0,1.5]$ each with a width of 0.05. The matrix $U_{ij}=P(w^{\rm{true}}_{i}|w^{\rm{rec}}_{j})$ models the probability that events reconstructed in the $w$ bin $j$ are in the true-$w$ bin $i$, which is calculated by Bayes' theorem according to

\begin{align}
	U_{ij} & = P(w^{\rm true}_{i} | w^{\rm rec}_{j}) \nonumber \\
	       & = P(w^{\rm rec}_{j}  | w^{\rm true}_{i}) \times P(w^{\rm true}_{i}) / P(w^{\rm rec}_{j}) \nonumber \\
	       & = P(w^{\rm rec}_{j}  | w^{\rm true}_{i}) \times P(w^{\rm true}_{i}) / \sum_{k} P(w^{\rm rec}_{j} | w^{\rm true}_{k}) P(w^{\rm true}_{k}). \label{eq:unfold}
\end{align}
Here, $P(w^{\rm{rec}}_{j}|w^{\rm{true}}_{i})$ is estimated with simulation. 
To avoid bias from the simulated signal, $P(w^{\rm true})$ is calculated using the reconstructed $w$ distribution on data as follows.
\begin{enumerate}
	\item $P(w^{\rm true}_{i})$ is assumed uniform ($P(w^{\rm true}_{i})=0.1$ for all bins).
	\item $U_{ij}$ is calculated by using Eq.(\ref{eq:unfold}).
	\item $P(w^{\rm true}_{i})$ is set to $ \sum_{j} U_{ij}(N^{\rm rec}_{j}-N^{\rm bg}_{j})/\sum_{ij} U_{ij}(N^{\rm rec}_{j}-N^{\rm bg}_{j}) $.
        \item Steps~2. and 3. are repeated 10~times, until $U_{ij}$ converges.
\end{enumerate}
The unfolding performance is validated with the simulation.

\subsection{Fitting method}

To determine $|V_{cb}|$, a binned maximum likelihood fit is performed using

\begin{align}
	\Delta \chi^{2} = -2 \ln \left(L\right) = 2\sum_{i} \left( N^{\rm{exp}}_{i} - N_{i} + N_{i} \ln \left(N_{i}/N^{\rm{exp}}_{i}\right) \right) + \prod_{i}\prod_{j} \Delta x_{i} W^{-1}_{ij} \Delta x_{j}, \label{ eq:likelihood }
\end{align}
where $i$ denotes the $w$ bin,  $N_{i}$ is the number of observed events in the $i$th bin, $x_{i}$ is a systematic parameter defined as the normalization uncertainty in the $i$th reconstructed-$w$ bin, $\Delta x_{i}$ is the deviation of the systematic parameters from the nominal value, and $W_{ij}$ is the covariance of the systematic parameters, modeled by multivariate Gaussians functions. Finally, $N^{\rm{exp}}_{i}$ is the expected yield in the $i$th bin, which is written as follows:

\begin{align}
	N^{\rm{exp}}_{i} \left(\Bz \rightarrow \Dstarm \ellp \nul \right)  &=  4 \epsilon_{i}  N_{\BB}  \left(1+f_{+0} \right)^{-1} \tau \left(\Bz \right) \mathcal{B}\left(\Dstarm \rightarrow \pim \Dzb \right)  \mathcal{B}\left(\Dzb \rightarrow \Kp \pim \right) \nonumber
\\	&   \left( 1 + \sum_{j} U_{ij}  \Delta x_{j} \right) \int^{w^{max}_i}_{w^{min}_i} dw \frac{d\Gamma}{dw} \left(\Bz \rightarrow \Dstarm \ellp \nul \right) , 
\end{align}
where $\epsilon_{i}$ is the signal reconstruction efficiency in the $i$th bin. The differential distribution is obtained using Eq.(\ref{ eq:dgdw }). 
In the fit there are two free parameters, $\eta_{\rm EW} F(1) |V_{cb}|$ and $\rho$, and ten nuisance parameters $\Delta x_{i}$.
The two-dimensional contour of $\eta_{EW} F(1) |V_{cb}|$ and $\rho$ is estimated by using a marginalized likelihood ~\cite{cite:pdgstat},

\begin{align}
	L_{\rm{marg}} = \frac{1}{J} \sum_{j=1}^{J} \exp \left( -\sum_{i} \left( N^{\rm{exp}}_{ij} - N_{i} + N_{i} \ln \left( N_{i}/N^{\rm{exp}}_{ij} \right) \right) \right), \label{ eq:likelihood_marg }
\end{align}
where $J=10000$ and $ N^{\rm{exp}}_{ij} $ is the expected yield in the $i$th bin with the $j$th set of nuisance parameters, which is generated following the covariance matrix. The fitter performance is validated with simplified simulated experiments.

\section{Systematic uncertainties} \label{ sec:sys }

\begin{figure}[!htb]
        \centering
        \includegraphics[width=8.0cm]{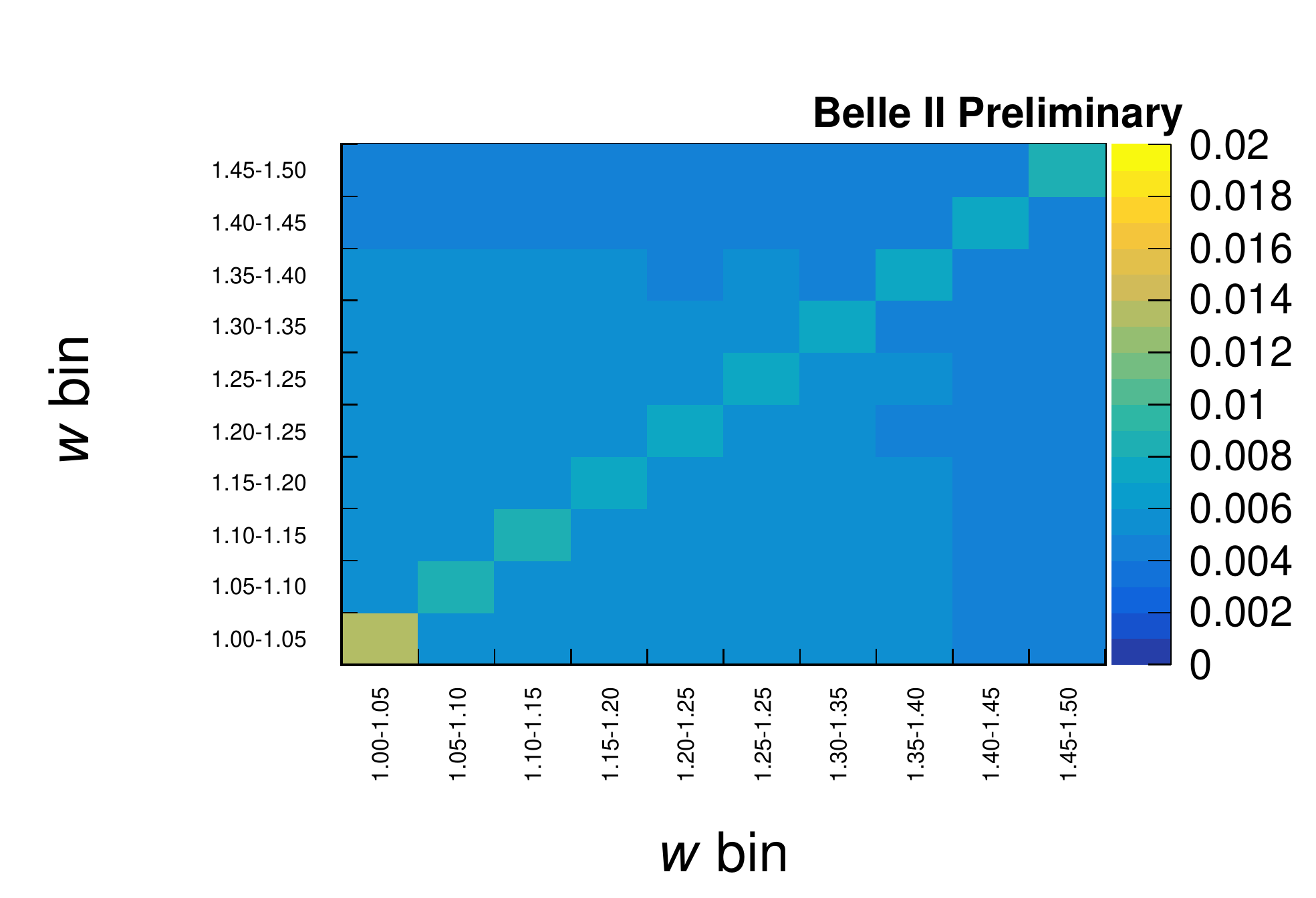}
	\caption{ Total covariance matrix for the $\eta_{EW} F(1) |V_{cb}|$ and $\rho^{2}$ measurement. The axes denote the $w$ bin intervals. }
  \label{ fig:cov }
\end{figure}

Systematic uncertainties are evaluated for several sources associated with the detector response, MC modeling, and physics inputs. For the branching ratio measurement, the systematic uncertainty of each source is propagated to the result based on Eq.(\ref{ eq:br0 })
and summarized in Table~\ref{ tab:sys_summary }. 
The $B_{\rm tag}$ reconstruction efficiency with the FEI algorithm is studied using $\B \rightarrow X \ell \nu$ decays and a systematic uncertainty of 3.9\% is assigned~\cite{cite:FEI}.
The tracking efficiency is studied with $\tau$ decays and the maximum data-simulation difference of 0.3\% is taken as systematic uncertainty for each track in the final state.
The reconstruction efficiency of the low momentum $\pim$ is studied by using $\Bz \rightarrow \pip \Dstarm (\Dstarm \rightarrow \pim \Dzb)$ decays. The data-MC ratio of the $\pi$ momentum distribution is evaluated relative to the high momentum distribution; a 3--4\% systematic uncertainty is assigned in each momentum bin, which is dominated by the statistical uncertainty of the control samples.
Electron and muon identification efficiencies and misidentification rates are studied by using $e^{+}e^{-}\rightarrow e^{+}e^{-}\ellp \ellm$, $e^{+}e^{-}\rightarrow e^{+}e^{-} (\gamma)$, $e^{+}e^{-}\rightarrow \mu^{+} \mu^{-} \gamma$, decays of $J/\psi$, \Dstar, $\tau$, and $K_{s}^{0}$. The lepton identification and misidentification uncertainties associated with the size of the control samples, background contamination, modeling of the fitting function, trigger, and the difference of the results across samples are evaluated as a functions of each lepton angle and the absolute value of the lepton momentum. These uncertainties are propagated to the branching fraction measurement resulting in a total 2.0\% systematic error.
The potential variations in the amount of background from $\B \rightarrow \Dstarstar \ell \nu$ decays, hadronic $\B$ decays and misreconstructed $\Dstar$ mesons are evaluated to propagate the uncertainty of the branching fraction of the background processes and of beam backgrounds resulting in a 1.2\% systematic uncertainty.
The number of produced $\BB$ pairs is estimated from the R2 distribution after a subtraction of the continuum background using off-resonance data. A systematic uncertainty of 2.9\% is assigned to account for the limited statistics of off-resonance data, operation conditions of the detector and accelerator including beam energy, and selection efficiencies.
A systematic uncertainty for the event-level selection is estimated to be 1.0\%, to cover the maximum data-simulation difference of the total energy in the electromagnetic calorimeter. 
The uncertainty from the limited size of simulated samples is estimated to be 1.8\%.
The following sources of systematic uncertainty are from external measurements: the ratio of the number of produced $\BpBm$ and $\BzBzb$ pairs (1.2\%), the ratio of the number of mixed and unmixed $\BzBzb$ (0.9\%), the branching fractions of $\Dstarm \rightarrow \pim \Dzb $ (0.7\%) and $\Dzb \rightarrow \Kp \pim$ (0.8\%), and form factors (0.1\%)~\cite{cite:pdg2020}.
The uncertainties from the various sources are assumed to be independent and the quadratic sum is taken as a total systematic uncertainty. For the measurement of $\eta_{EW} F(1) |V_{cb}|$ and $\rho^{2}$, the effect of the systematic uncertainty is included in the likelihood calculation (the second term in Eq.(\ref{ eq:likelihood })) with the covariance matrix

\begin{align}
	W_{ij} = \sum_{k}  \frac{\left(N_{i}^{k}-\mu_{i} \right) \left( N_{j}^{k}-\mu_{j} \right)}{\mu_{i}\mu_{j}}. 
\end{align}
Here, $k$ runs over the sources of uncertainties, $\mu_{i}$ is the mean of the expected yield in the $i$th $w$ bin, $N_{i}^{k}$ is the variation of the expected yield in the $i$th bin for the $k$th source of uncertainties. Figure~\ref{ fig:cov } shows the estimated covariance matrix.

\begin{table}[!htb]
	\begin{center}
		\caption{ Summary of fractional systematic uncertainties on the branching ratio. }
		\label{ tab:sys_summary }
		\begin{tabular}{lccccccccccc}
			\hline
			\hline
			Systematic sources & Relative uncertainty (\%)  \\
			\hline
			FEI efficiency                    &  3.9    \\
			Low momentum $\pi$ efficiency     &  4.1    \\  
			Tracking efficiency               &  0.9    \\
			Lepton particle identification    &  2.0    \\
			Background                        &  1.2    \\
			$N_{\BB}$                         &  2.9    \\
			$f_{+0}$                          &  1.2    \\
			Number of mixed \BB               &  0.9    \\
			$\mathcal{B}\left(\Dstarm \rightarrow \pim \Dzb \right)$          &  0.7    \\
 			$\mathcal{B}\left(\Dzb \rightarrow \Kp \pim \right)$              &  0.8    \\
			ECL energy                        &  1.0    \\
			Form factor                       &  0.1    \\
			MC sample size                    &  1.8    \\
			\hline
			Total                             &  7.3    \\
			\hline
			\hline
		\end{tabular}
	\end{center}
\end{table}

\section{Results and conclusion}

\begin{figure}[!htb]
        \centering
        \includegraphics[width=10.0cm]{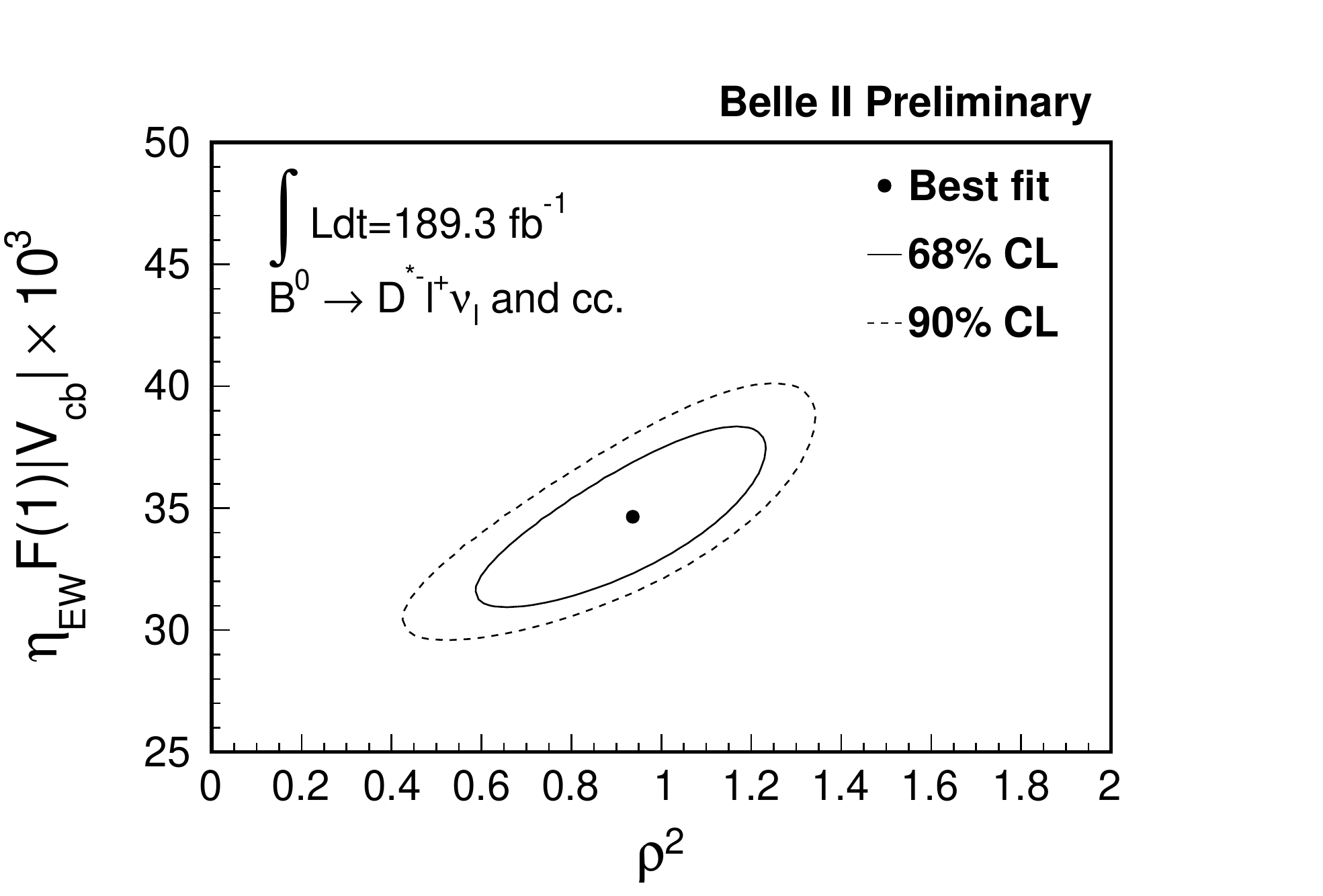}
\caption{ Two dimensional probability contours for $\eta_{EW} F(1) |V_{cb}|$ and $\rho^{2}$ at the 68\% (solid) and 90\% (dashed) confidence level. The best fit point is also shown. }
  \label{ fig:result }
\end{figure}

\begin{figure}[!htb]
        \centering
        \includegraphics[width=10cm]{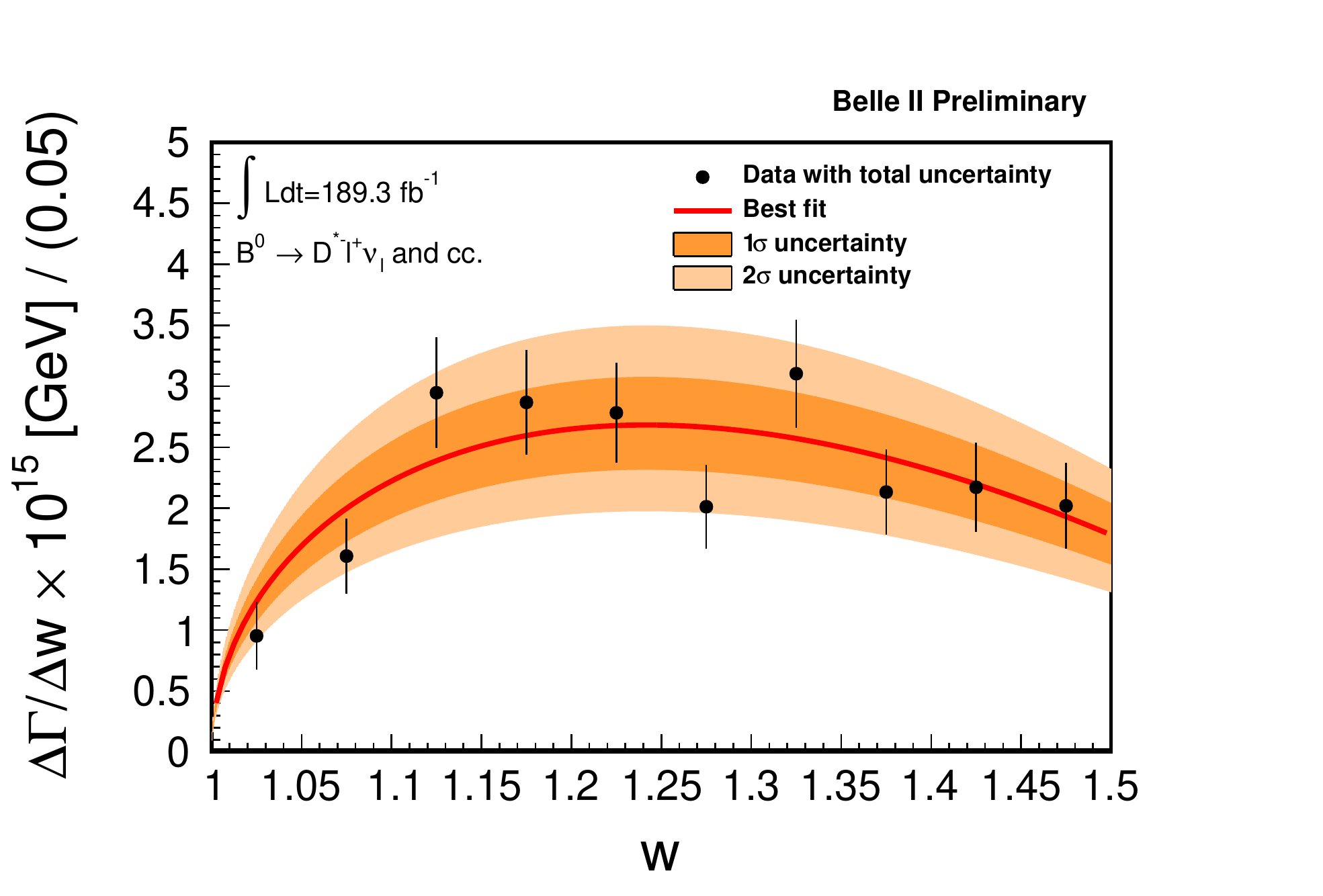}
\caption{ Observed $d\Gamma(\Bz \rightarrow \Dstar \ell \nu)/dw$ distribution with the best fit function and one and two standard-deviation bands overlaid. }
  \label{ fig:fitresult }
\end{figure}

The result for the branching fraction is

\begin{align}
	\mathcal{B} \left(\Bz \rightarrow \Dstarm \ellp \nul \right)  &=  \left( 5.27 \pm 0.22~\left(\rm{stat}\right) \pm 0.38~\left(\rm{syst}\right) \right) \%
\end{align}
while the results for $|V_{cb}|$ are

\begin{align}
	\eta_{EW} F(1) |V_{cb}| \times 10^{3}  &= 34.6 \pm 1.8 ~\left(\rm{stat}\right) \pm 1.7~\left(\rm{syst}\right) 
	\\                                                                                           
	\rho^{2}   &=   0.94   \pm 0.18 ~\left(\rm{stat}\right) \pm 0.11 ~\left(\rm{syst}\right) .
\end{align}
The two-dimensional probability contours for $\eta_{EW} F(1) |V_{cb}|$ and $\rho^{2}$ are shown in Fig.~\ref{ fig:result }. The observed $\ \Delta \Gamma / \Delta w$ values are shown in Fig.~\ref{ fig:fitresult } with the best fit function overlaid. The reduced $\chi^{2}$ of the fit is 1.6 with p-value of 40.7$\,\%$, which is estimated by simulation. Under the assumption that $\eta_{\rm EW}=1.00662 \pm 0.00016$ and 
$F(1) = 0.906 \pm 0.004 ~\rm{(stat)} \pm 0.012~\rm{(syst)}$~\cite{cite:qcd2}, we obtain $|V_{cb}| \times 10^{3}=37.9 \pm 2.7 $. The results are consistent with the world averages of
$\mathcal{B} (\Bz \rightarrow \Dstarm \ellp \nul ) = (5.06 \pm 0.12) \%$ and
$\eta_{EW} F(1) |V_{cb}| \times 10^{3} =  35.27 \pm 0.38 $ based on exclusive $\B \rightarrow \Dstar \ell \nul$ decays
within one standard deviation~\cite{HFLAV:2019otj}.

\section{Acknowledgement}
These acknowledgements are not to be interpreted as an endorsement of any statement made by any of our institutes, funding agencies, governments, or their representatives.

We thank the SuperKEKB team for delivering high-luminosity collisions;
the KEK cryogenics group for the efficient operation of the detector solenoid magnet;
the KEK computer group and the NII for on-site computing support and SINET6 network support;
and the raw-data centers at BNL, DESY, GridKa, IN2P3, INFN, and the University of Victoria for offsite computing support.

\bibliography{belle2_short}
\bibliographystyle{belle2-note}

\end{document}